\documentclass[twocolumn,showpacs,preprintnumbers,amsmath,amssymb]{revtex4}

\usepackage{graphicx}
\usepackage{dcolumn}
\usepackage{bm}
\usepackage[usenames]{color}
\usepackage{amssymb,amsfonts,amsmath}
\usepackage{bbold}
\usepackage{epstopdf}
\usepackage{lpic}
\usepackage{capt-of}

\newcommand{\comment}[1]{}
\newcommand{\NT}[1]{{\color{black}#1}}

\begin{document}

\title{Multi-shell model of ion-induced nucleic acid condensation}

\author{Igor~S.~Tolokh}
\affiliation{Department of Computer Science, Virginia Tech, Blacksburg, VA 24061, USA}
\author{Aleksander~Drozdetski}
\affiliation{Department of Physics, Virginia Tech, Blacksburg, VA 24061, USA}
\author{Lois~Pollack}
\affiliation{Cornell University, School of Applied and Engineering Physics, Ithaca, NY 14853-3501, USA}
\author{Nathan~A.~Baker}
\affiliation{Computational and Statistical Analytics Division, Pacific Northwest National Laboratory, Richland, WA 99352, USA}
\author{Alexey~V.~Onufriev}
\affiliation{Departments of Computer Science and Physics, Virginia Tech, Blacksburg, VA 24061, USA}

\date{\today}

\begin{abstract}
We present a semi-quantitative model of condensation of short nucleic acid (NA) 
duplexes induced by tri-valent cobalt(III) hexammine (CoHex) ions.  
The model is based on partitioning of bound counterion distribution 
around single
NA duplex into ``external" and ``internal" ion binding shells
distinguished by the proximity to duplex helical axis. 
In the aggregated phase the shells overlap, which leads to  
significantly increased attraction of CoHex ions in these overlaps
with the neighboring duplexes. 
The duplex aggregation free energy is decomposed into attractive and repulsive components
\NT{in such a way that they can be} represented by simple analytical expressions 
\NT{with parameters derived from molecular dynamic (MD) simulations and numerical
solutions of Poisson equation.}
\NT{The short-range interactions described by the attractive term depend on 
the fractions of bound ions in the overlapping shells and 
affinity of CoHex to the ``external" shell of nearly neutralized duplex.} 
\NT{The repulsive components of the free energy 
are duplex configurational entropy loss upon the aggregation 
and the electrostatic repulsion of the duplexes that remains after 
neutralization by bound CoHex ions.}
The estimates of the aggregation free energy are consistent
with the experimental range of NA duplex condensation propensities,
including the unusually poor condensation of RNA structures 
and subtle sequence effects upon DNA condensation.
The model predicts that, in contrast to DNA, RNA duplexes may condense into
tighter packed aggregates with a higher degree of duplex neutralization. 
The model also predicts that longer NA fragments will condense more readily than shorter ones.
The ability of this model to explain experimentally observed trends in NA condensation, 
lends support to proposed NA condensation picture based on  
the multivalent ``ion binding shells".
\end{abstract}


\maketitle
\clearpage

\section{INTRODUCTION}

Condensation of highly charged DNA and RNA molecules by cationic agents is biologically 
important for processes such as packaging of genetic material inside living cells 
and viruses \cite{Luger1997,Belyi2006,Marenduzzo2009,Borodavka2012,Garmann2014}, 
compactization and delivery of small interfering RNA molecules for gene silencing 
\cite{Agrawal2003} and gene therapy \cite{Mansoori2014}.
In aqueous solution, DNA condensation requires cations with charges of +3$e$ or higher 
\cite{Widom1980,Bloomfield1991,Bloomfield1996}; 
e.g., trivalent cobalt(III) hexammine (CoHex), trivalent spermidine or 
tetravalent spermine can generally condense DNA at room temperatures, 
while divalent cations cannot.

Several decades of experimental and theoretical studies of DNA condensation have 
resulted in a general physical picture wherein the major contribution to 
the effective attraction is due to electrostatic interactions 
\cite{Gosule1976,Gosule1978,Wilson1979,Guldbrand1986,Braunlin1987,RauParsegian1992,Lyubartsev1995,Rouzina1996,Rouzina1997,Bloomfield1997,Levin1999,Kornyshev1999,Shklovskii1999,Nguyen2000,Matulis2000,Kornyshev2007,Huang2008,Loth2009,Kanduc2010}.
Theoretical models have previously been developed to clarify the details of 
nucleic acid (NA) interactions, starting from the models of interacting 
uniformly charged cylinders immersed in an implicit ionic bath 
\cite{Oosawa1968,Brenner1974,Stigter1977,Naji2004}, 
to nucleic acid models with more realistic helical geometries of 
molecular charge distributions 
\cite{Edwards1994,Kornyshev1997,Kornyshev1998,Allahyarov2004,Tan2005,Tan2006b,Kornyshev2007,Kanduc2009a}.
Counterion electrostatic treatments range from simple mean-field descriptions to strong coupling models 
\cite{Rouzina1996,Nguyen2000a,Levin2002,Naji2004,Naji2005,Kanduc2009a,Kanduc2010} 
with the more sophisticated strong coupling models able to reproduce NA-NA attraction.
It has been shown that mean-field description of the counterion atmosphere always leads
to repulsion between the oppositely charged cylinders \cite{Guldbrand1986,Kanduc2010} --
correlations between counterions is a necessary condition for 
the attraction \cite{Oosawa1968,Levin2002,Deserno2003,Naji2005}.
\NT{All-atom explicit solvent molecular dynamics (MD) simulations of short DNA fragments 
demonstrated the existence of short-range attractive forces between B-form DNA 
duplexes at sufficient degree of duplex neutralization by multivalent ions
\cite{Luan2008,Dai2008,Wu2015,Yoo2016}.}

Despite progress in understanding NA-NA interactions, an atomic-level mechanism 
of multivalent ion-induced nucleic acid condensation has not yet been fully established
and some recent experimental data are difficult to rationalize within 
the accepted models. 
For example, the condensation propensity of double-stranded (ds) RNA 
in the presence of trivalent CoHex ion was recently found to be much smaller 
than for the equivalent sequence of dsDNA \cite{Li2011,Tolokh2014}. 
In other words, dsRNA 
helices resist CoHex-induced condensation under conditions where the DNA duplexes 
readily condense. 
The unexpected findings are still very recent, with only a limited number of 
theoretical models attempting to rationalize it so far. For 
example, a recent extension \cite{Kornyshev2013} of an earlier 
model \cite{Kornyshev1999} of DNA condensation suggests 
that the striking difference between RNA and DNA condensation stems
from the differences in intrinsic helical parameters of the duplexes.  
However, it has recently been shown experimentally that
significant differences in condensation propensity can arise in some NA
duplexes without significant differences in their helical parameters \cite{Tolokh2014}. 
Nevertheless, the model \cite{Kornyshev2013} emphasizes critical role of counterion
distributions in NA condensation; 
some of the distributions of bound trivalent counterions assumed by the model 
are consistent with the latest all-atom  molecular dynamics (MD) simulations
\cite{Tolokh2014}, but some are very different, pointing to the critical 
need to take these atomistic details into account for quantitative and 
even qualitative agreement with recent condensation experiments.   

Although it is possible to count the number of excess ions around nucleic acids, using 
techniques like buffer equilibration and atomic emission spectroscopy (BE-AES) \cite{Bai2007} 
or an indicator dye \cite{Grilley2009}, only absolutely calibrated small angle 
X-ray scattering (SAXS) experiments can both count the number of ions \cite{Pabit2010}
and reveal their locations \cite{Pabit2009_MIE,Meisburger2015}. 
Even so, only the ensemble average distribution can be detected.
At this point, only atomistic simulations that treat the solvent (water and ions) 
explicitly can provide several key details of counterion binding and distributions 
around nucleic acids, which are otherwise near impossible to obtain
experimentally. Among these critical details is thermodynamic characterization 
of CoHex binding to different regions of NA, which, as we shall see, is necessary 
for a quantitative description of condensation. However, 
at this point experiment can only provide average, aggregate 
binding affinity to NA, but this is not the key quantity of interest for 
condensation \cite{Tolokh2014}. Atomistic simulations can bridge the gap 
with experiment in that respect and provide estimates of the counterion 
distributions and affinities to various loci in the NA duplexes. In what 
follows we show that such estimates become valuable in building 
a quantitative picture of NA condensation -- picture that can be verified
experimentally.  

\subsection{The ``shells'' model of CoHex distributions around NA duplexes}

\begin{figure}[t]
\vspace*{0.1in}
\begin{center}
\includegraphics[scale=0.16] {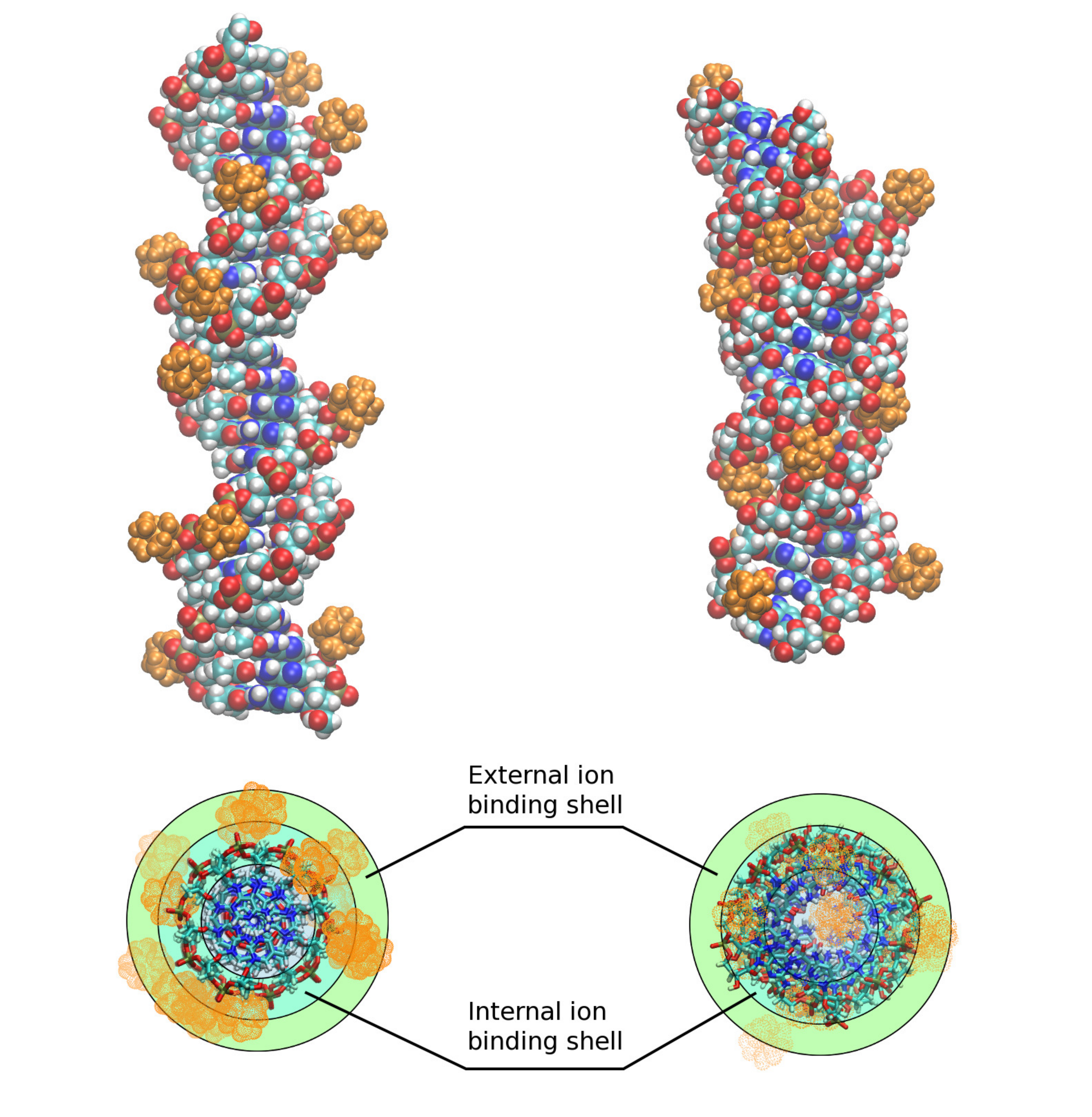}
\caption{The ``shells" of counterion distribution around nucleic acids duplexes \cite{Tolokh2014}.
Upper panel: representative snapshots of B-form dsDNA (left) 
and A-form dsRNA (right) structures with bound CoHex ions (orange). 
Lower panel: the ``external" and ``internal" cylindrical ion binding shells  
around B-DNA (left) and A-RNA (right). 
Most of the ions ($\sim$ 2/3) shown in the upper panel are bound in the ``external" shell 
of the B-DNA, and the ``internal" shell of the A-RNA.}  
\label{fig:DNA_RNA}
\end{center}
\end{figure}

Recent all-atom MD simulations \cite{Tolokh2014} of 
\NT{
short (25 base pairs) DNA, RNA and DNA:RNA hybrid duplexes with CoHex counterions
}
show that, under near neutralization conditions 
necessary for NA condensation ($\sim$ 90\% of NA charge neutralized by bound counterions), 
the majority ($\sim$ 2/3) of neutralizing CoHex ions bound to B-form DNA are
localized at the external surface of the phosphate groups in the cylindrical layer
12 to 16 \AA\ from the helical axis, Fig.~\ref{fig:DNA_RNA}. This layer 
is defined as the ``external'' ion binding shell \cite{Tolokh2014}.
The smaller fraction of bound CoHex ions around B-DNA ($\sim$ 1/3) are localized 
in the major groove at distances 7 to 12 \AA\ from 
the helical axis -- the corresponding cylindrical layer is
defined as the ``internal'' ion binding shell \cite{Tolokh2014}.
In contrast to B-DNA, the majority of CoHex ions ($\sim$ 2/3) bound to A-form
RNA are localized within the RNA major groove,
in the ``internal'' ion binding shell, Fig.~\ref{fig:DNA_RNA}.
\NT{
This substantial difference in CoHex distributions is explained by 
the major difference in the electrostatic potential of
B-DNA and A-RNA \cite{Chin1999,Tolokh2014}. The potential in the major groove 
of A-RNA is about 10 kcal$/$mol$/|e|$ lower than on the rest of the RNA 
surface  accessible to CoHex, or anywhere on the surface of
B-DNA.
The resulting much stronger attraction of trivalent CoHex ions overwhelms 
the ion-ion repulsion, leading to the qualitatively different pictures in 
CoHex binding.
}

An analysis of the simulated CoHex ion distributions around 25-bp
DNA, RNA and DNA:RNA hybrid duplexes of the equivalent mixed sequence \cite{Pabit2009b} 
and around homopolymeric poly(dA):poly(dT) DNA 
led to the following two observations \cite{Tolokh2014}.
\NT{
    First, at the near-neutralizing conditions necessary for NA condensation,
the fractions of CoHex ions in the ``external'' ion binding shells of 
the four NA duplexes correlate with measured condensation propensities of these duplexes
\cite{Tolokh2014}.
    Second, the ``external'' CoHex binding shells of NA molecules 
are the only shells that overlap substantially at the inter-axial duplex separation
corresponding to the separations observed in CoHex condensed (aggregated) DNA phases
(about 28 \AA\ \cite{RauParsegian1992,Qiu2013}).
}

These two observations constitute a basis for the ``overlapping ion binding shell'' 
mechanism of NA condensation which was proposed
in our previous paper \cite{Tolokh2014}. 
According to the mechanism, the fraction of neutralizing multivalent ions
bound in the NA ``external'' ion binding shell (and not the total number 
of bound ions) is a key parameter for understanding
multivalent ion-induced nucleic acid condensation. This fraction reflects 
a complex interplay between various structural and sequence features of NA
helices, and its ions are responsible for most of the attractive interaction
between the helices. The proposed mechanism was rationalized by simple and 
robust electrostatic arguments, and is in excellent qualitative agreement 
with experimental condensation propensities of various NA duplexes.
\NT{
The mechanism 
was later illustrated by explicit calculations of the potential of mean force (PMF) 
between two adjacent NA duplexes \cite{Wu2015}.
}
However, no quantitative model of the duplex condensation phenomenon 
has been presented so far; such a
model would be the best illustration for the proposed new mechanism and  
could lead to novel predictions. 

	In this work, we develop a semi-quantitative model of CoHex-induced NA duplex 
condensation that is based on the ``overlapping ion binding shell'' mechanism.
\NT{
The model provides a quantitative relationship between key characteristics of bound 
multivalent counterion distributions around NA molecules and
the free energy changes upon NA duplex aggregation.
}
We show that the estimated free energy changes for different NA duplexes
correlate well with the experimental condensation propensities of these duplexes
and explain observed differences in DNA and RNA condensation, 
and make several predictions.

\section{The multi-shell model of ion-mediated NA-NA interaction}

We consider a transition between a dilute aqueous solution of 
relatively short nucleic acid duplexes (compared to the DNA persistence 
length of $\sim$150 base pairs) and their condensed (aggregated) phase
represented by a bundle of parallel, hexagonally packed NA molecules \cite{Qiu2013} 
with a distance between neighbors smaller than their length.
The solution phase contains a certain amount of trivalent CoHex
counterions sufficient to neutralize all the duplexes.
Part of these ions are bound to the duplexes due to Manning-Oosawa
condensation \cite{Manning1969,Oosawa1971,Manning1978,Shklovskii1999a}.
The fraction of NA duplex charge neutralized by bound CoHex ions, $\Theta$,
in the solution and aggregated phases is considered to be the same,
although the aggregated phase as a whole is assumed to be neutral.

The aggregated phase is stabilized by short-range attractive forces between 
the duplexes that originate mostly from the electrostatic interactions
of multivalent counterions bound to one duplex with the ``correlation holes''
\cite{Shklovskii1999a,Levin1999,Nguyen2000a,Nguyen2000,Levin2002}
on the charged surface of another duplex covered by a layer 
of bound multivalent counterions. I.e., these interactions have a significant 
non-mean-field component \cite{Oosawa1968,Naji2004,Loth2009,Kanduc2010}
due to correlations between multivalent counterions bound to different
neighboring duplexes in the aggregate.

Our goal is to estimate the free energy difference between the aggregated 
and solution phases of NA duplexes, $\Delta G_{aggr}$, as a function of $\Theta$.
The latter can be used as a convenient proxy for bulk CoHex concentration
in the solution phase. 

The multivalent ion-induced aggregation free energy
can be represented as a sum of three additive components:
\begin{equation}
\Delta G_{aggr} = \Delta G_{attr} + \Delta G_{el-rep} - T\Delta S_{conf} \, .
\label{dGagr}
\end{equation}
The first two terms in Eq.~\ref{dGagr} describe the changes in the electrostatic
interactions upon the transition between the solution and aggregated phases.
These interactions are decomposed into the short-range net attractive term, $\Delta G_{attr}$,
and the net repulsive term, $\Delta G_{el-rep}$ that 
describes the (residual) repulsion 
between the duplexes almost neutralized by bound counterions.
The last term, $T\Delta S_{conf}$, represents the loss of
duplex configurational entropy (translational and rotational) upon the aggregation. 

\NT{
The decomposition of the electrostatic contributions to $\Delta G_{aggr}$ in
Eq.~\ref{dGagr} into net attractive and net repulsive terms is not unique.
Both terms include contributions from the interactions of multivalent ions 
bound to one duplex with the bare charges of adjacent duplex, 
which are favourable for the aggregation,
and contributions from the interactions of these same 
ions with the ions bound to the adjacent duplex,
which oppose the aggregation. Due to ion-ion correlations this latter type
of the interactions between bound multivalent ions is difficult to estimate.
However, grouping these interactions between bound multivalent ions
into the contributions where the ion-ion correlations play a substantial role
and where they can be neglected allows us to estimate the net attractive
$\Delta G_{attr}$ and net repulsive $\Delta G_{el-rep}$ electrostatic terms 
through simple analytical expressions.
}

The last term in Eq.~\ref{dGagr}, $-T\Delta S_{conf}$, can be estimated using
a simple coarse-grained approach. This term is usually neglected for long DNA
molecules \cite{Nguyen2000}, but, as we shall see later, it can contribute
appreciably to the destabilization of the aggregated phase in the case of
relatively short 25-bp NA duplexes.


Additional underlying assumptions of our model for the aggregation free energy,  
physical considerations that justify them, and details of how each term 
in Eq.~\ref{dGagr} is calculated are presented in ``METHODOLOGICAL DETAILS". 
Derivations of the components of Eq.~\ref{dGagr} are described below.

\subsection{Short-range attractive component, $\Delta G_{attr}$, of 
the ion-mediated duplex-duplex interactions}

\NT{
The multivalent ion-mediated short-range attractive forces between the duplexes 
in the aggregated phase arise from the interactions of 
the multivalent counterions bound to the surface of one duplex with
all the charges on the neighboring duplex (including its bound counterions).
These are the same interactions that determine the distribution of
counterions around NA duplexes in the solution phase. They are strong
when the counterion is inside the duplex ion binding shell and relatively weak
when outside \cite{Ramanathan1983,Tan2005}, and can be characterized
by the ion binding affinity to nearly neutralized duplex.
In the case of multivalent counterions these interactions can have 
a significant non-mean-field component \cite{Naji2004,Loth2009,Kanduc2010}
due to ion-ion correlations which reduce the ion-ion repulsion.

In our decomposition of the aggregation free energy, Eq.~\ref{dGagr}, we consider 
that these attractive interactions contribute to $\Delta G_{attr}$ 
when CoHex ion bound to one duplex enters the ``external'' ion binding shell 
of another duplex \cite{Ray1994}, i.e. when the ``external'' binding shells 
of the two duplexes overlap upon aggregation. 
When the bound CoHex ion is outside the ion binding shell of a neighboring duplex
we no longer consider its interaction with that duplex 
contributing to $\Delta G_{attr}$. In this case the bound ion
can be treated as part of the averaged neutralizing background
\cite{Ramanathan1983,Stigter1995,Deserno2000,OShaughnessy2005}
that screens the bare duplex charge and reduces the mutual electrostatic
repulsion between the duplexes described by $\Delta G_{el-rep}$.
The proposed decomposition assumes that the non-mean-field component 
of the ion interactions with the counterions bound to the neighboring duplex
can be significant and accounted for in $\Delta G_{attr}$ when
the ion enters the ``external'' ion binding shell of the neighboring duplex
but is small and can be neglected when the ion is outside the ion binding shell.
}

\NT{
Once the duplexes approach each other upon aggregation, CoHex ions bound 
in the ``external" ion binding shell of one duplex enter the ``external'' 
shell of another duplex. The additional interaction energy of each of these 
ions in the ``external" ion binding shell of the neighboring duplex 
is essentially the binding energy for the additional CoHex
ions in the ``external'' shell of the duplex. 
Assuming that minimal changes in the CoHex distributions occur when 
the two duplexes approach each other, this binding energy averaged 
over the volume the shell can be approximated by CoHex binding affinity, 
$\mu_a$, to the ``external" ion binding shell of an isolated NA duplex.
This quantity varies with the degree of duplex neutralization $\Theta$,
but for a narrow range of $\Theta$ (at near neutralizing conditions,
$0.9 \le \Theta \le 1.0$)
that is of interest to us here, this dependence is small and can be neglected. 
Ion affinity $\mu_a$ reflects the balance between the attraction to the bare 
NA duplex charges and the repulsion from all other bound ions around the duplex
and, therefore, depends on the ion charge, $Ze$. It also absorbs 
the contribution from the ion-ion correlations which increases with 
the ion valency $Z$.
}

The above assumptions about the contributions to $\Delta G_{attr}$
result in a short-range attractive component of the interactions between 
the two neighboring NA duplexes that is proportional to two quantities: 
(1) the average number of CoHex ions in the overlapping region 
of the ion binding shells of these duplexes, $\Delta N_s$,
and (2) 
the binding affinity $\mu_a$ of 
CoHex ion in the ``external'' ion binding shell of NA duplex 
at near neutralizing conditions.

For the hexagonal packing of the duplexes in the aggregated phase \cite{Qiu2013},
the total short-range attractive term in the decomposition of the aggregation 
free energy, Eq.~\ref{dGagr}, per duplex can therefore be estimated as
\begin{equation}
\Delta G_{attr} = 3 \mu_a \times \Delta N_s \, ,
\label{eq:dGattr}
\end{equation}
where the factor of 3 accounts for the half of the attractive interactions 
with the six nearest neighbors in the hexagonally packed aggregate.

To estimate $\mu_a$ and $\Delta N_s$ we will use the equilibrium properties
of CoHex distribution around a single NA duplex in the solution phase
obtained from the MD simulation.

\subsubsection{CoHex binding affinity $\mu_a$.}

The binding affinity of an ion to the ion binding shell of a polymer
can be defined as a difference of the excess chemical potentials of this ion
in the binding shell and in the bulk. In the case of the ``external" ion 
binding shell of NA duplex where the excluded volume effects for CoHex ions 
are negligible this difference can be estimated as
\begin{equation}
 \mu_a = - k_B T\ln \left(\rho_s/\rho_b\right) \, ,
\label{mua}
\end{equation}
where $\rho_b$ and $\rho_s$ are the number densities of CoHex ions in the bulk
and in the ``external" ion binding shell of the NA duplex, respectively.
A more detailed derivation of this equation is presented
in ``METHODOLOGICAL DETAILS".

Because of the finite size of the simulation box used in all-atom MD simulations
\cite{Tolokh2014} and the absence of a monovalent salt screening in the simulations,
the CoHex concentration at the simulation box boundary, $\rho_B$,
is not equal to the bulk value $\rho_b$ required in Eq.~\ref{mua} for estimating $\mu_a$.
To account for the difference, we introduce a long-range correction
to $\mu_a$:
\begin{equation}
\mu_a = - k_B T\ln \left(\rho_s /\rho_B\right) + Ze\varphi(r_B) \, ,
\label{mua_corr}
\end{equation}
where $Ze\varphi(r_B)$ is the energy of the CoHex ion charge, $Ze$, 
in the electrostatic potential of the NA duplex and its bound ions, $\varphi(r_B)$, 
at the system boundary ($r_B=31$ \AA).
Without monovalent ions present, $\varphi(r)$ can be estimated
as the electrostatic potential of a uniformly charged non-conducting rod of length $H$
with a linear charge density $\lambda$ in a solvent with dielectric constant $\varepsilon$,
\begin{equation}
\varphi(r)=\frac{2\lambda}{\varepsilon}\ln\left(\frac{H/2+\sqrt{(H/2)^2+r^2}}{r}\right) \, .
\label{eq:phi}
\end{equation}
Here $\lambda$ corresponds to the linear charge density of 25-bp NA duplex
(charge $Q_{na}=-48e$, length $H$) re-normalized (scaled down) by the charge of 
all bound CoHex ions ($N_0 Ze$) within the outer boundary of 
the ``external" ion binding shell,
$\lambda=(Q_{na} + N_0 Ze)/H = Q_{na}(1-\Theta_0)/H$,
where $\Theta_0 = N_0 Ze/|Q_{na}|$ is a degree of duplex neutralization by
all ($N_0$) bound CoHex ions determined from the results of MD simulations \cite{Tolokh2014}.
The potential $\varphi(r)$ is zero at infinity.
The values of the potential estimated for the NA duplexes of interest 
at $r_B = 31$ \AA\ and dielectric constant $\varepsilon=78.5$ 
are relatively small, ranging from -0.7 to -1.1 $k_BT/e$.

\subsubsection{Number of bound CoHex ions, $\Delta N_s$, in the overlapping volume 
of ion binding shells.}

\begin{figure}[h!]
\vspace*{0.1in}
\begin{center}
\includegraphics[scale=0.7] {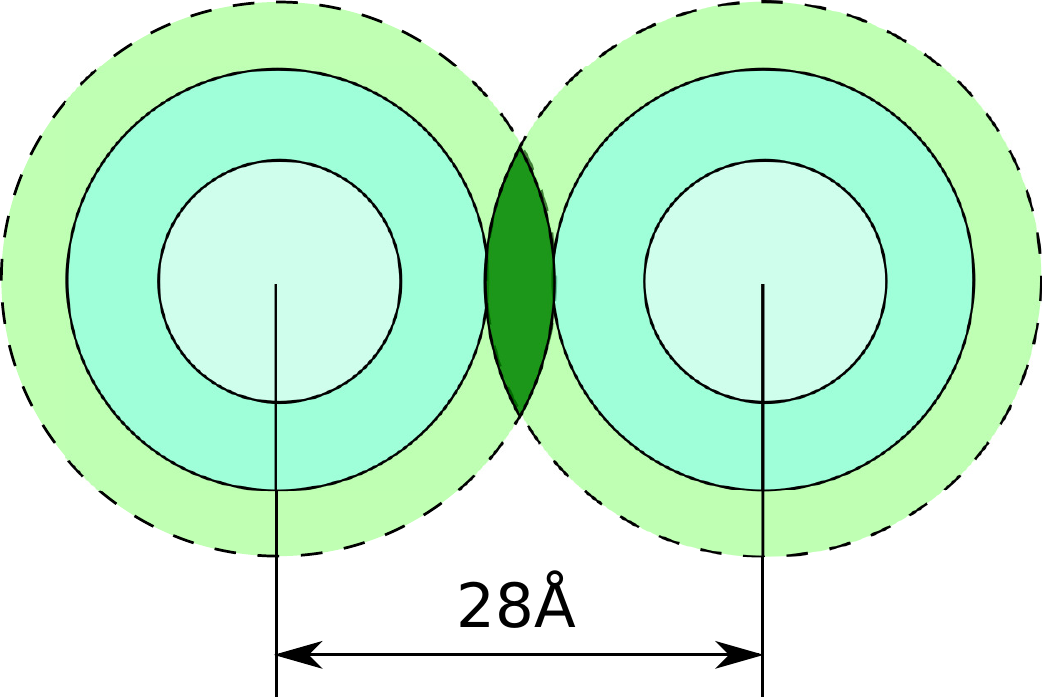}
\caption{Horizontal cross section of the overlap of the ``external"
CoHex ion binding shells of two NA duplexes at the inter-axial separation $d=28$ \AA.
The overlap region (volume) is indicated by dark green color.
CoHex ions in this volume element, which is about 80 \AA\ high 
(length of the NA duplex),
strongly interact with both duplexes creating the effective short-range attraction
between NA molecules. These ions are excluded from the estimation of the
effective duplex-duplex electrostatic repulsion in Eq.~\ref{eq:correction}.}
\label{fig:overlap}
\end{center}
\end{figure}

Fig.~\ref{fig:overlap} shows a schematic of the ``internal" and ``external"
ion binding shells around two parallel duplexes at the typical
duplex-duplex separation of 28 \AA\
in the DNA aggregate \cite{RauParsegian1992,Qiu2008,Qiu2013}.
The shells overlap volume is completely defined by geometries of the shells
and mutual orientation of the duplexes. 
In the hexagonal packing of the duplexes in the aggregated phase \cite{Qiu2013}
considered in our model, the duplexes are parallel to each other
and their ends are assumed at the same height.

Under the assumption of minimal changes in the CoHex distributions
when the duplexes approach each other
the number of ions $\Delta N_s$ in the overlap volume
of the two ``external" ion binding shells \cite{SI} can be estimated as
\begin{equation}
\Delta N_s = 2 \rho_s \Delta V_s = 2\, \frac{N_s}{V_s}\Delta V_s \, ,
\label{eq:dN_s}
\end{equation}
where $V_s$ and $N_s$ are the volume and the average number of CoHex ions 
bound in the ``external" ion binding shell, respectively.

Since we are interested in exploring the dependence of the aggregation free 
energy $\Delta G_{aggr}$ on the degree of NA duplex neutralization $\Theta$
at near neutralizing conditions when $\Theta$ is close to its value $\Theta_0$
observed in MD simulations, we need to estimate how $N_s$ changes 
with $\Theta$. Assuming the fractions of the bound CoHex ions 
in each of the ion binding shell are insensitive to the small changes
in the total number of bound CoHex ions, $N$, we can write
\begin{equation}
N_s = f_s N = f_s \frac{|Q_{na}|}{Ze} \Theta \, ,
\label{eq:N_s}
\end{equation}
where $f_s=N_s^0/N_0$ is the fraction of CoHex ions bound in the ``external" 
shell, $N_s^0$ and $N_0$ are the simulation results values \cite{Tolokh2014}.

\NT{
The above results allow one to rewrite Eq.~\ref{eq:dGattr} as
\begin{equation}
\Delta G_{attr}(\Theta) = 6 \mu_a \frac{\Delta V_s}{V_s} \frac{|Q_{na}|}{Ze} f_s \Theta \, .
\label{eq:dGattr2}
\end{equation}
Note that for long duplexes, $\mu_a$ is 
independent of the duplex length $H$, and so $\Delta G_{attr}$ in
Eq.~\ref{eq:dGattr2} depends on $H$ only through the bare duplex charge 
$Q_{na}= \lambda_{na} H$,
where $\lambda_{na}$ is the (constant) 
linear charge density of the unscreened NA duplex.
}

\subsection{Repulsive electrostatic component $\Delta G_{el-rep}$ of 
the ion-mediated duplex-duplex interactions}

Our decomposition of the electrostatic interactions between two NA duplexes
upon ion-mediated duplex aggregation onto
the attractive and repulsive parts is based on the separation of the bound
CoHex ions into those which are in the ion binding shell overlap
volume $\Delta V_s$, Fig.~\ref{fig:overlap}, and the rest of the bound ions.
The interactions of the former ones with the neighboring duplex
and all of its bound counterions have a significant ion-ion correlation contribution
and constitute the attractive part of the aggregation energy
$\Delta G_{attr}$ described above.
The interactions of rest of the bound ions with the neighboring duplexes and their
bound counterions can be treated at the mean-field level. In this case
these counterions can be considered as a neutralizing background which
uniformly reduces the bare charge of each duplex.
This approximation allows us to use continuum electrostatics in estimating
$\Delta G_{el-rep}$.

As discussed above, the NA duplex neutralization by bound CoHex is described
by a degree of neutralization $\Theta$ which, according to Manning counterion
condensation theory \cite{Manning1978} applied to NA molecules interacting
with trivalent counterions, reaches $\sim 0.92$ level. Approximately the same
level of duplex neutralization by CoHex ions ($\Theta_0 \sim 0.88 - 0.92$) 
is observed in all-atom MD simulations \cite{Tolokh2014}.

As in the case of $\Delta G_{attr}$, 
we consider the repulsive interactions between the duplexes at zero monovalent salt
condition, neglecting a small monovalent ion screening in the NA duplex condensation
experiments \cite{Li2011,Tolokh2014}.
Indeed, for spermine$^{4+}$, which is similar to CoHex in its condensing
power, condensation of 150-bp DNA fragments was shown
to be unaffected by monovalent ion concentrations of 20mM
and lower if DNA monomer concentration ([P]) is higher than 0.4 mM \cite{Raspaud1998}.
In the condensation experiments we discuss \cite{Li2011,Tolokh2014},
NA monomer concentration is about 2 mM.

\NT{
Additional considerations of why
the influence of 20mM of NaCl on CoHex induced duplex condensation can
be neglected here include:
1) the interactions of bound CoHex ions with NA molecules are virtually unaffected 
by monovalent salt concentration below 40 mM \cite{Braunlin1987} and
2) CoHex affinity $\mu_a$ to NA duplex is affected by a small amount of monovalent salt
through the change of long-range potential in much the same way
as the long-range repulsive duplex-duplex interactions.
Since each of these two monovalent salt screening effects on
$\Delta G_{attr} \sim \mu_a$ and $\Delta G_{el-rep}$
are small and opposite, we assume that the combined salt screening effect on
the duplex-duplex attractive and repulsive terms is negligibly small and estimate
these terms without considering the monovalent screening.
However, at physiological monovalent salt concentrations ($\sim$ 100 mM),
the difference could increase; the drop in the attraction due to reduction of CoHex affinity
may be substantially larger than the drop in the repulsive term, contributing to the onset
of NA duplex aggregation by monovalent ions \cite{Raspaud1998}.
}

The above approximations (no ion-ion correlations and
zero monovalent salt condition) allow us to estimate 
the duplex-duplex repulsion 
as interaction between the neighboring NA duplexes with
their charges uniformly re-normalized by bound counterions, with
the duplexes being in continuum dielectric without free ions.
The interactions beyond the nearest neighbors in the hexagonal
packing are neglected (assumed to be screened out by monovalent ions
or by unbound CoHex ions in the net neutral duplex aggregated phase).

Within the linear response electrostatics in a continuum dielectric,
the interaction between two duplexes with the charges re-normalized
uniformly by a factor $\xi$ is equivalent to scaling 
the interaction between bare duplexes, $\Delta G_{el-na}$, by $\xi^2$. 

To estimate $\xi$ we notice that according to the specific way we decompose 
the electrostatic interactions, all CoHex ions bound to one NA duplex
($N \sim \Theta$), 
except those ($\Delta N_s/2$) in the shell overlapping volume,
participate in the duplex charge re-normalization.
Taking into account Eqs.~\ref{eq:dN_s} and \ref{eq:N_s},
the charge fraction of these $\Delta N_s /2$ ions
that has to be excluded from the re-normalization is
\begin{equation}
\Lambda = \frac{\Delta N_s}{2} \frac{Ze}{|Q_{na}|} =
\frac{\Delta V_s}{V_s} f_s \Theta \, .
 \label{eq:eta}
\end{equation}
This fraction reduces the duplex screening and changes the duplex
charge re-normalization coefficient from $\xi = 1-\Theta$
to $\xi = 1-\Theta + \Lambda$.
The resulting estimate for the duplex repulsive contribution to the aggregation
free energy at the hexagonal packing of the aggregated phase
can be written as
\begin{eqnarray}
\Delta G_{el-rep}(\Theta) &=& 
3 \Delta G_{el-na} (1 - \Theta  + \Lambda)^2  \label{eq:correction}  \\
&=& 3 \Delta G_{el-na} \left(1 - \Theta \left[1-\frac{\Delta V_s}{V_s} f_s \right] \right)^2 ,  \nonumber
\end{eqnarray}
where $\Delta G_{el-na}$ is the electrostatic interaction between
the two bare duplexes in a dielectric continuum which
can be estimated numerically, within the Poisson framework.
Here we neglect
the interactions with more distant neighbors in the hexagonal packing
taking into account that they should be scaled by a much smaller
factor ($\xi^2 =(1-\Theta)^2$ at $\Lambda=0$) and assuming the overall neutrality
of the aggregated phase due to unbound diffuse ions.

\NT{
Note that the re-normalization coefficient $\xi$ is independent of 
the duplex length $H$ and so the repulsion term $\Delta G_{el-rep}$ 
in Eq.~\ref{eq:correction}
depends on $H$ only through the bare duplexes repulsion $\Delta G_{el-na}$.
For long duplexes $\Delta G_{el-na}\sim \lambda_{na}^2 H$ 
leading to $\Delta G_{el-rep}$ linear with $H$, similar to
$\Delta G_{attr}$. 
}

We estimate $\Delta G_{el-na}$ by solving the Poisson equation (PE) for
the two parallel unscreened duplexes and averaging the interaction energies
over 12 different mutual duplex orientations.
The details of this estimation are presented in ``METHODOLOGICAL DETAILS".

\subsection{Configurational entropy loss upon duplex association}

        The change in the NA duplex configuration entropy upon duplex aggregation,
$\Delta S_{conf}$, is caused by restrictions of the
duplex translational and rotational
motion in the aggregated phase compared to a free motion in a dilute solution phase.
For long NA molecules this entropic contribution
to the aggregation free energy is usually neglected compared to other repulsive and
attractive contributions \cite{Nguyen2000}.
For short 25 base pair NA duplexes, however, the contribution of $T\Delta S_{conf}$
can be substantial and, as we will see, comparable with the electrostatic repulsion
between the duplexes in the aggregated phase.

\begin{figure}[h!]
\begin{center}
\includegraphics[scale=0.44] {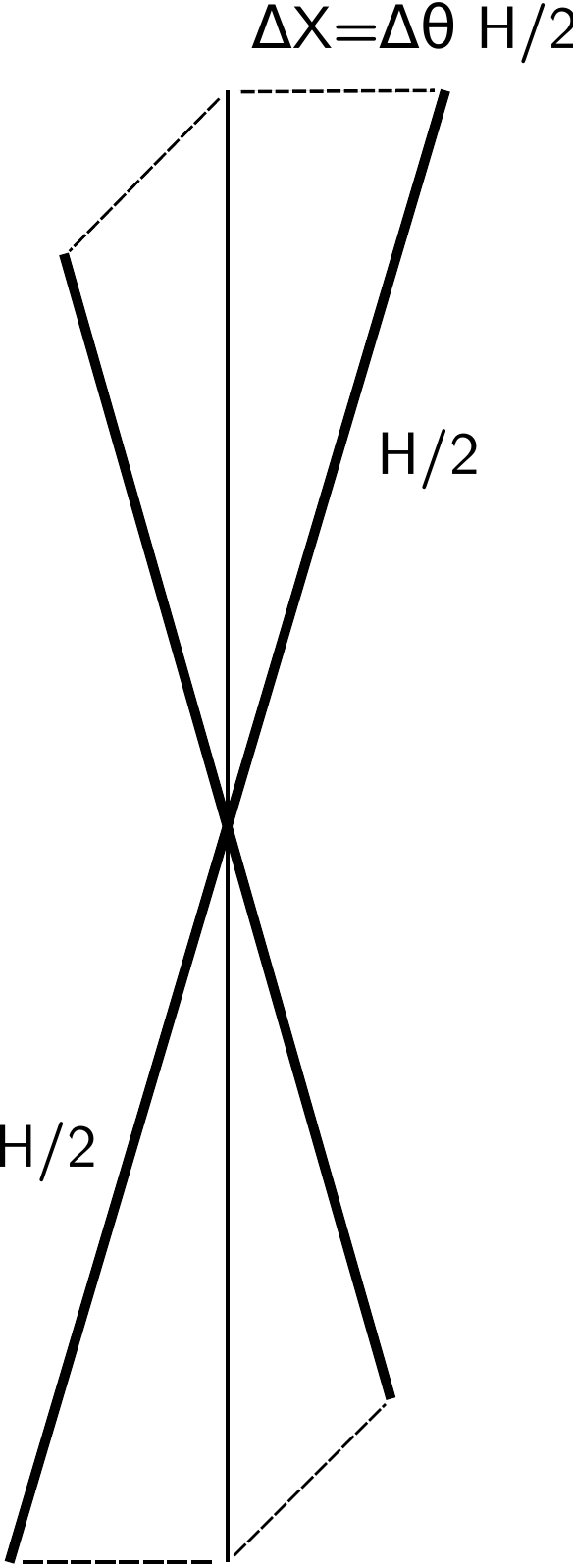}
\caption{Schematic of a rigid rod rotation about the two axes used in the
estimate of the rotational entropy change upon aggregation.
}
\label{fig:rod}
\end{center}
\end{figure}

Short NA duplexes of the length smaller than the NA persistence length ($\sim 150$-bp 
for dsDNA) can be considered as rigid rods. 
We estimate their configuration entropy change,
$\Delta S_{conf} = \Delta S_{tran} + \Delta S_{rot}$,
by following the approach described in Ref.~\cite{Swanson2004},
\begin{equation}
\Delta S_{conf} =  k_B \ln(c\Delta X \Delta Y \Delta Z) + k_B \ln( \Delta X \Delta Y / \pi H^2) \, .
\label{eq:entrop}
\end{equation}
Here, the first term corresponds to the change in the duplex translational entropy $S_{tran}$
due to reduction of volume available for translational motion
of the duplex center of mass
from $1/c$ in the solution phase to $\Delta X \Delta Y \Delta Z$ in the aggregated phase,
where $c$ is a NA duplex concentration in the solution.
The second term reflects the loss of the rotational entropy $\Delta S_{rot}$ in two of the three
duplex rotational degrees of freedom upon aggregation;
the corresponding reduction of the available rotational phase space being
$\Delta \theta_1 \Delta \theta_2 /4\pi$ where
$\Delta \theta_{1(2)} \approx \Delta X(Y) /(H/2)$, Fig. \ref{fig:rod}.
In our estimations we use $\Delta X =\Delta Y = (d - 2a)/2$ 
where $d$ is inter-axial duplex separation and $a$ is the radius of the NA duplex
estimated as $a=11$ \AA. At $d=28$ \AA\ corresponding to duplex separation
observed in CoHex condensed DNA phases \cite{RauParsegian1992,Qiu2013},
$\Delta X =\Delta Y = 3$ \AA\ -- a half-width of the nucleic acid 
effective attraction energy well \cite{Kornyshev2013} at 5$k_BT$ above the minimum.
For simplicity we use the same value for $\Delta Z$ ($\Delta Z = 3$ \AA),
which describes the vertical misalignment of the ends of parallel duplexes
in the hexagonal bundle.

\NT{
From Eq.~\ref{eq:entrop} it follows that only $\Delta S_{rot}$ 
part of the duplex configuration 
entropy change depends on the duplex length $H$ -- this part scales 
logarithmically with $H$, increasing its contribution
to the aggregation energy for longer duplexes.
}

\subsection{Attractive and repulsive interactions at the ``internal-external" 
ion binding shell overlaps}

So far, we considered the simplest case of two parallel NA duplexes at such separations
in the aggregated phase that only their ``external" ion binding shells overlap;
we approximated the additional binding free energy for the CoHex ion bound in
the ``external" ion binding shell of one duplex and entering the ``external" ion binding
shell of the neighboring duplex by the CoHex binding affinity $\mu_a$ to the ``external" shell.
This affinity, defined as a difference of the CoHex excess chemical potentials,
Eq.~\ref{mua_def}, reflects only the change of CoHex ion interactions with its environment
and does not depend on CoHex configurational entropy.
Since this entropy in either of the two ``external" ion binding shells
of the two neighboring duplexes is same, the above approximation 
is reasonable.

\begin{figure}[h!]\vspace*{0.1in}
\begin{center}
\includegraphics[scale=0.7] {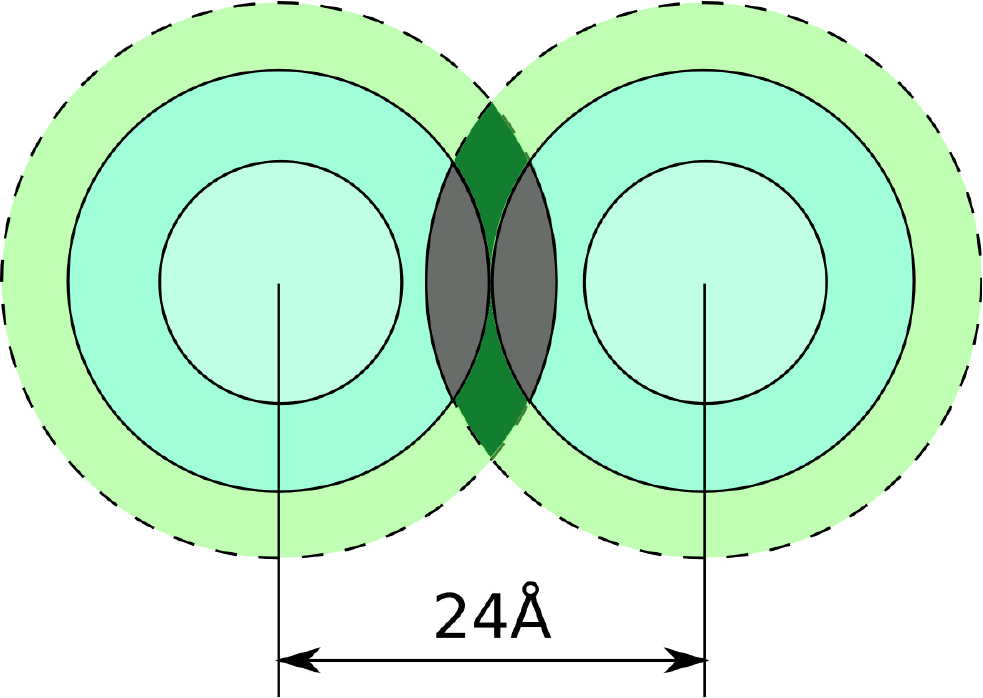}
\caption{Schematic of the overlaps of the ``external" and ``internal"
ion binding shells of two NA duplexes at the inter-axial separation $d=24$ \AA.
The regions of the ``external-internal" shells overlap are gray. Dark green
indicates ``external-external" shells overlap as in Fig.~\ref{fig:overlap} }
\label{fig:overlap2}
\end{center}
\end{figure}

If the duplex-duplex separations in the aggregated phase are smaller than 28 \AA,
in addition to the ``external-external" shell overlaps the overlaps of
the ``external" and ``internal" ion binding shells of the adjacent duplexes occur,
Fig.~\ref{fig:overlap2}. Below we will argue that the attractive contribution
$\Delta G_{attr}$
to the aggregation energy can still be estimated by Eq.~\ref{eq:dGattr}
where the quantity $\Delta N_s$ is replaced by $\Delta N$, the sum of the numbers of
bound CoHex ions in both overlaps.

In equilibrium, the chemical potential of CoHex ions in the ``external" and ``internal"
ion binding shells is the same. Any difference in the excess chemical potentials
of the ions in these shells (and, thus, in the ion binding affinities) is compensated by
the difference in ion configurational entropies within these shell. The same consideration
can be applied to the estimation of the free energy gain for the CoHex ion bound
in the ``external" shell of one duplex once entering the ``internal" shell of adjacent
duplex when these shells overlap. Whatever correction to $\mu_a$
might be applied due to the difference in the CoHex excess chemical potentials in these
shells, it should be compensated by the original difference in the ion configurational entropies
in these shells.
Therefore, the contribution of each CoHex ion in the ``external-internal" shell overlaps
to $\Delta G_{attr}$ can still be approximated by the same value of $\mu_a$ as in the case
of the CoHex ions in the ``external-external" shell overlaps.
This approximation allows us to use Eq.~\ref{eq:dGattr} for
the close inter-duplex separation, Fig.~\ref{fig:overlap2}, as well,
but now $\Delta N_s$ becomes $\Delta N$, 
the sum of the numbers of ions in two overlapping shell regions,
``external-external'' and ``external-internal" (combined gray and
dark green regions in Fig.~\ref{fig:overlap2}).
\NT{
This quantity can be estimated as 
\begin{eqnarray}
\Delta N &=& 2 \rho_s \Delta V_s + (\rho_i + \rho_s) \Delta V_{is} \nonumber \\
&=& \left[
\frac{f_{i}}{V_{i}} \Delta V_{is} + \frac{f_s}{V_s}(\Delta V_{is} + 2\Delta V_s)
\right]  \frac{|Q_{na}|}{Ze} \Theta \, ,
\label{eq:dNsum}
\end{eqnarray}
where $\Delta V_{is}$ is the volume of the ``external-internal" overlap, 
$\rho_i$ and $f_i$ are the number density and the fraction 
of CoHex ions bound in the ``internal" ion binding shell of the volume $V_{i}$.
Similar to $f_s$, the fraction $f_i$ is
defined as $f_i=N_i^0/N_0$ where $N_i^0$ is the number of CoHex ions in 
the ``internal" shell determined from the all-atom MD simulations \cite{Tolokh2014}.
Estimations of $N_i^0$, $N_0$ are in Table~\ref{tbl:NAR}, while 
$\Delta V_{is}$ and $\Delta V_s$ are 
presented in the Supplemental Material \cite{SI}.

Using Eq.~\ref{eq:dNsum}, the attractive contribution to the aggregation free energy,
Eq.~\ref{eq:dGattr2}, can be rewritten as
\begin{equation}
\Delta G_{attr}(\Theta) = 3 \mu_a \left[
\frac{f_{i}}{V_{i}} \Delta V_{is} + \frac{f_s}{V_s}(\Delta V_{is} + 2\Delta V_s)
\right]
\frac{|Q_{na}|}{Ze} \Theta \, .
\label{eq:dGattr3}
\end{equation}
}

We do not consider the ``internal-internal" shell overlaps, as these could only
occur at such short inter-duplex separations where steric repulsion
would become prohibitively large.

\NT{
In the case of ``external-internal" ion binding shell overlap,
the quantity $\Delta N$, Eq.~\ref{eq:dNsum}, has to be used instead of $\Delta N_s$ 
in determining the correction $\Lambda$, Eq.~\ref{eq:eta},
to the duplex charge re-normalization coefficient $\xi = 1-\Theta + \Lambda$.
The latter one can now be written as
\begin{equation}
\xi = 1-\Theta + \frac{1}{2} \left[
\frac{f_{i}}{V_{i}} \Delta V_{is} + \frac{f_s}{V_s}(\Delta V_{is} + 2\Delta V_s)
\right] \Theta \, ,
\label{eq:xisum}
\end{equation}
resulting in the following duplex-duplex repulsive term in the aggregation
free energy,
\begin{eqnarray}
\Delta G_{el-rep}(\Theta) = 3 \Delta G_{el-na} \times \hspace{4cm} \label{eq:rep3} \\
\times \left(1-\Theta
\left[1 - \frac{1}{2}\left(\frac{f_{i}}{V_{i}}\Delta V_{is} + 
\frac{f_s}{V_s}(\Delta V_{is} + 2\Delta V_s) \right) \right] \right) ^2 . \nonumber
\end{eqnarray}

As in the case of the duplex separations in the aggregated phase when only 
the ``external" ion binding shells overlap, for long duplexes 
the attractive and repulsive contributions, Eqs.~\ref{eq:dGattr3} and \ref{eq:rep3}, 
depend linearly on the duplex length $H$ through the bare duplex charge, 
$Q_{na}= \lambda_{na} H$, and bare duplexes repulsion,
$\Delta G_{el-na}\sim \lambda_{na}^2 H$, respectively. The entropic
contribution, Eq.~\ref{eq:entrop}, remains unchanged and scales
logarithmically with $H$.


Note that Eqs.~\ref{eq:dGattr3} and \ref{eq:rep3} reduce to  simpler 
Eqs.~\ref{eq:dGattr2} and \ref{eq:correction}  when the ``external-internal"
overlap volume $\Delta V_{is}$ vanishes for the inter-axial duplex separations
$d \ge r_i + r_s = 28$ \AA~~ ($r_i + r_s$ is 
the sum of the outer radii of the ``internal" and
``external" ion binding shells).
Since the typical inter-axial duplex separation in the CoHex induced DNA aggregates
is 28 \AA\ \cite{RauParsegian1992,Qiu2008,Qiu2013}, for this and larger separations
we can use more simple Eqs.~\ref{eq:dGattr2} and \ref{eq:correction}.
}

\section{Application of the Model  and Discussion}

\subsection{The stability of nucleic acid aggregates.}

We apply our semi-quantitative model, Eq.~\ref{dGagr}, where the terms are 
determined by Eqs.~\ref{eq:dGattr3},\ref{eq:rep3} and \ref{eq:entrop},
to characterize CoHex induced aggregation of four 25-bp long nucleic acid duplexes
previously simulated and studied experimentally in \cite{Tolokh2014};
these are DNA, RNA and DNA:RNA hybrid duplexes of
the equivalent mixed sequence \cite{Andersen2008,Pabit2009b} and homopolymeric
poly(dA):poly(dT) DNA duplex.

The model presented here allows us to estimate the aggregation free energies of these
duplexes, $\Delta G_{aggr}$, as functions of the degree of duplex neutralization
by bound CoHex ions, $\Theta$, and compare them with the duplex condensation propensities
observed in \cite{Tolokh2014}.
The aggregation begins when $\Theta$, which depends on the bulk CoHex concentration
in the solution phase, reaches the level at which the NA aggregated phase is more stable
than the solution phase, i.e. when $\Delta G_{aggr}<0$.

\begin{table}[ht!]
\begin{center}
 \vspace*{.01in}
\captionof{table}{Estimated changes of the free energy,
$\Delta G_{aggr} = \Delta G_{attr} + \Delta G_{el-rep} - T\Delta S_{conf}$,
upon CoHex induced 25-bp NA duplex aggregation under experimentally relevant conditions
(CoHex concentration corresponds to 94\% of duplex charge neutralization,
NA duplex concentrations in the solution phase is 40 $\mu$M).
All energy components are in $k_BT$ units per one duplex, $T=300$ K.
The calculations assume 28 \AA\ inter-axial duplex separation \cite{RauParsegian1992,Qiu2013}
in the hexagonal aggregated phase that
corresponds to ``external-external" CoHex binding shells overlap shown in
Fig.~\ref{fig:overlap}.
}
\label{tbl:28A}
 \begin{tabular}{@{\vrule height 10.5pt depth4pt  width0pt}|c|c|c|c|c|}
\hline
                        & dA:dT &    DNA    &   HYB  &   RNA  \\
\hline
CoHex binding affinity,
             $\mu_a$    & -7.66  &  -7.57   & -7.34  & -5.79  \\
\hline
Fraction of ``externally"  bound & & & & \\ 
   ions, $f_s$          &  0.67  &  0.59   &   0.24  &  0.14 \\
\hline
Number of ions in single & & & & \\
overlap, $\Delta N_s$   &  2.41  &   2.12  &   0.88  &  0.49  \\
\hline
$\Delta G_{attr}$       & -55.5  &  -48.2  &  -18.6  &  -8.4  \\
\hline
$\Delta G_{el-rep}$     &  27.6  &   24.0  &   11.5   &  8.7  \\
\hline
$-T\Delta S_{conf}$     &  22.2  &   22.2  &   21.9  &   21.9  \\
\hline
$\Delta G_{aggr}$       &  -5.7  &   -2.0  &   14.8  &   22.2  \\
\hline
\end{tabular}
 \vspace*{.05in}
\end{center}
\end{table}
%

\begin{figure}[ht!]
\vspace*{0.4in}
\begin{center}
\includegraphics[scale=0.34] {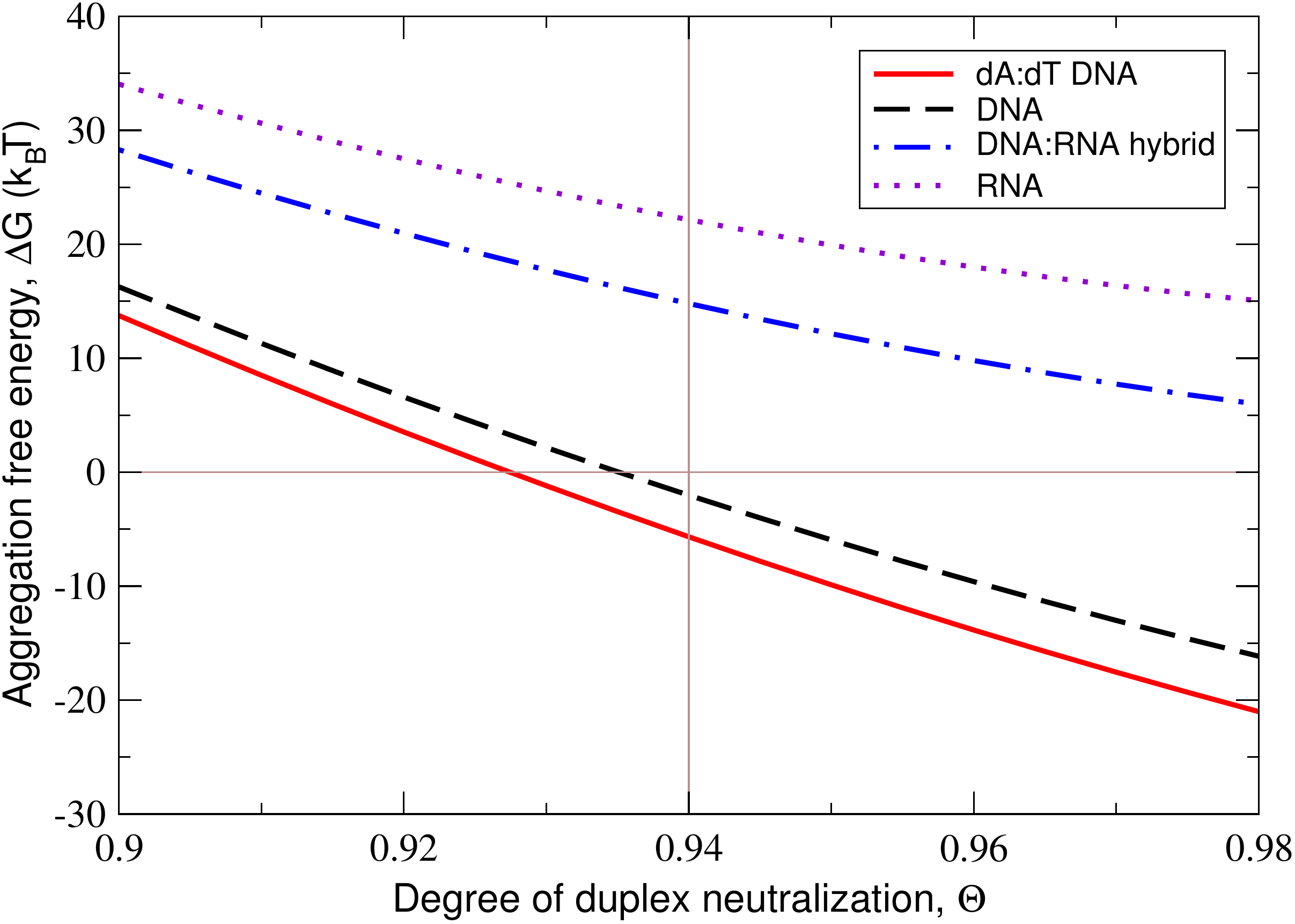}
\caption{Predicted aggregation free energy $\Delta G_{aggr}$ for 
the four 25-bp NA duplexes as a function of the degree of duplex charge 
neutralization $\Theta$ by bound CoHex ions.
The calculations assume 28 \AA\ inter-axial duplex separation in the hexagonal 
aggregated phase and 40 $\mu$M duplex concentration in the solution phase.
The data in the Table~\ref{tbl:28A} correspond to $\Theta=0.94$ (thin vertical line).  
}
\label{fig:dG}
\end{center}
\end{figure}


Our main results for $\Delta G_{aggr}$ and its components at the typical 28 \AA\ 
inter-axial duplex separation in the hexagonal aggregated phase 
are summarized in Table~\ref{tbl:28A} and Fig.~\ref{fig:dG}. 
The latter shows $\Delta G_{aggr}$ as a function of $\Theta$.
In Table~\ref{tbl:28A}, the attractive, repulsive and configurational entropy 
components of $\Delta G_{aggr}$ for the four NA duplexes are compared
for the same degree of duplex neutralization $\Theta=0.94$, at which
the aggregated phase for the two considered DNA duplexes is more stable
than the solution phase, $\Delta G_{aggr}<0$.
The main factors determining the CoHex ion-mediated duplex attraction,
CoHex binding affinity to the ``external" ion binding shell $\mu_a$
and the fraction of bound CoHex ions in this shell $f_s$, 
are presented as well. 

The predicted aggregation free energies suggest that at the same degree of
duplex neutralization the DNA aggregates are more stable than 
the DNA:RNA hybrid or RNA aggregates (see Fig.~\ref{fig:dG}).
Among the two DNA molecules, the mixed sequence DNA duplex requires 
higher degree of neutralization by CoHex ions 
for the beginning of the aggregation, $\Delta G_{aggr}<0$.
Provided that the CoHex affinities $\mu_a$ for both DNA duplexes 
are roughly the same (see Table~\ref{tbl:28A}), 
the higher $\Theta$ at $\Delta G_{aggr}=0$ for the mixed sequence DNA 
translates into the higher bulk CoHex concentration 
in the solution phase at which mixed sequence DNA duplexes condense. 
At roughly the same $\mu_a$ for both DNA duplexes, 
the major difference in the values of the attractive components stems 
from the difference in the fractions of bound CoHex ions in 
the ``external" ion binding shells $f_s$ of these duplexes,
0.67 for the homopolymeric DNA and 0.59 for the mixed sequence DNA,
leading to different values of the number of CoHex ions $\Delta N_s$
in the overlaps of these shells for two adjacent NA duplexes
in the aggregated phase.
%

Comparing the DNA:RNA hybrid and RNA duplexes, the differences in both
$\mu_a$ and $f_s$ (and, therefore, $\Delta N_s$) contribute to a smaller value 
of the attractive component in $\Delta G_{aggr}$ for the RNA duplex. 
The same conclusion is valid when comparing DNA and RNA duplexes. 
However, the major part of the difference in $\Delta G_{attr}$ for 
DNA and RNA duplexes arises from a drastic difference in $f_s$,
0.59 and 0.14, respectively (see Table~\ref{tbl:28A}), 
leading to four fold difference in the number of CoHex ions $\Delta N_s$
in the shell overlaps.

\subsection{Dependence of the NA aggregation free energy on inter-axial 
duplex separation}

\begin{figure}[h!]
\begin{center}
\includegraphics[scale=0.24] {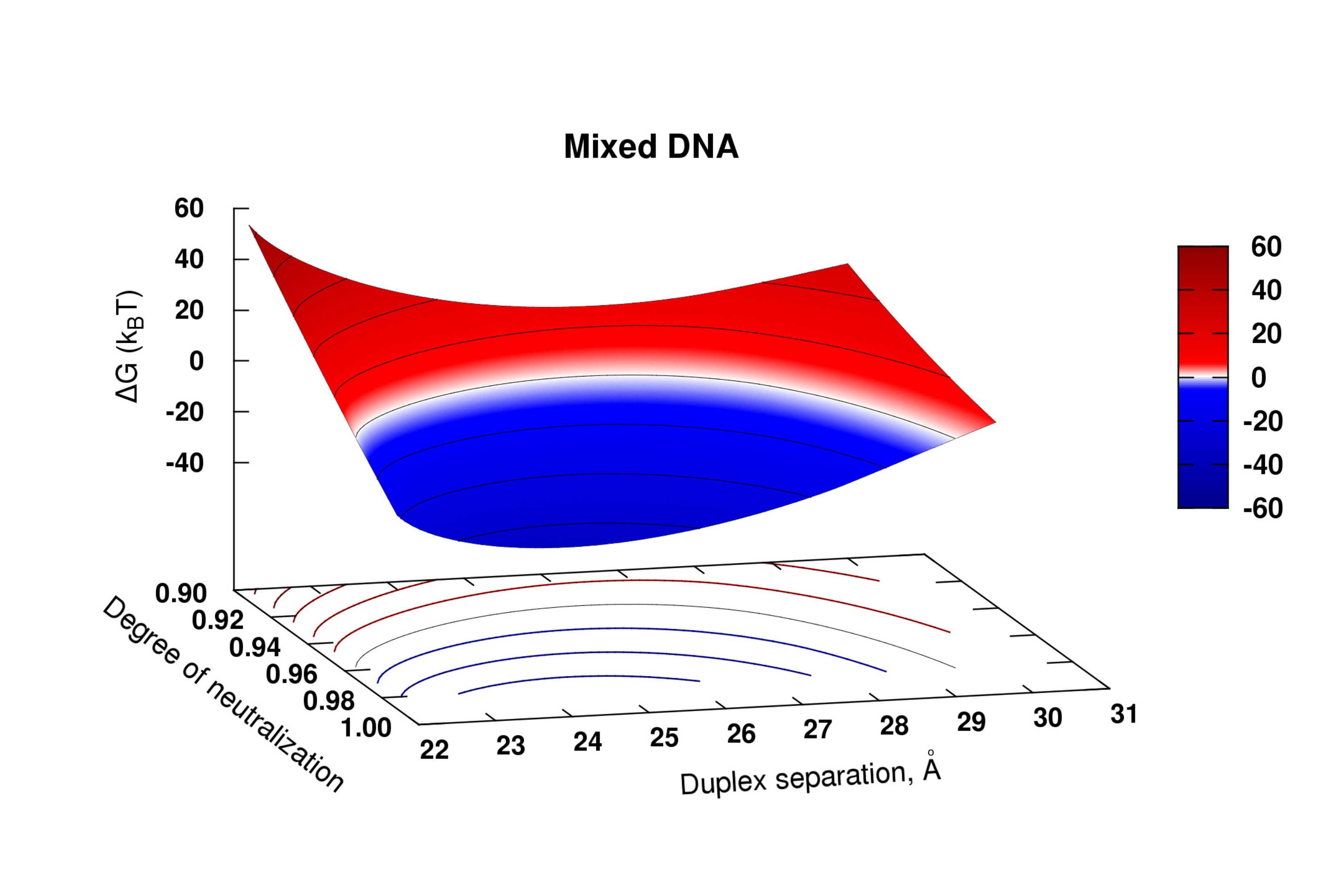}
\includegraphics[scale=0.24] {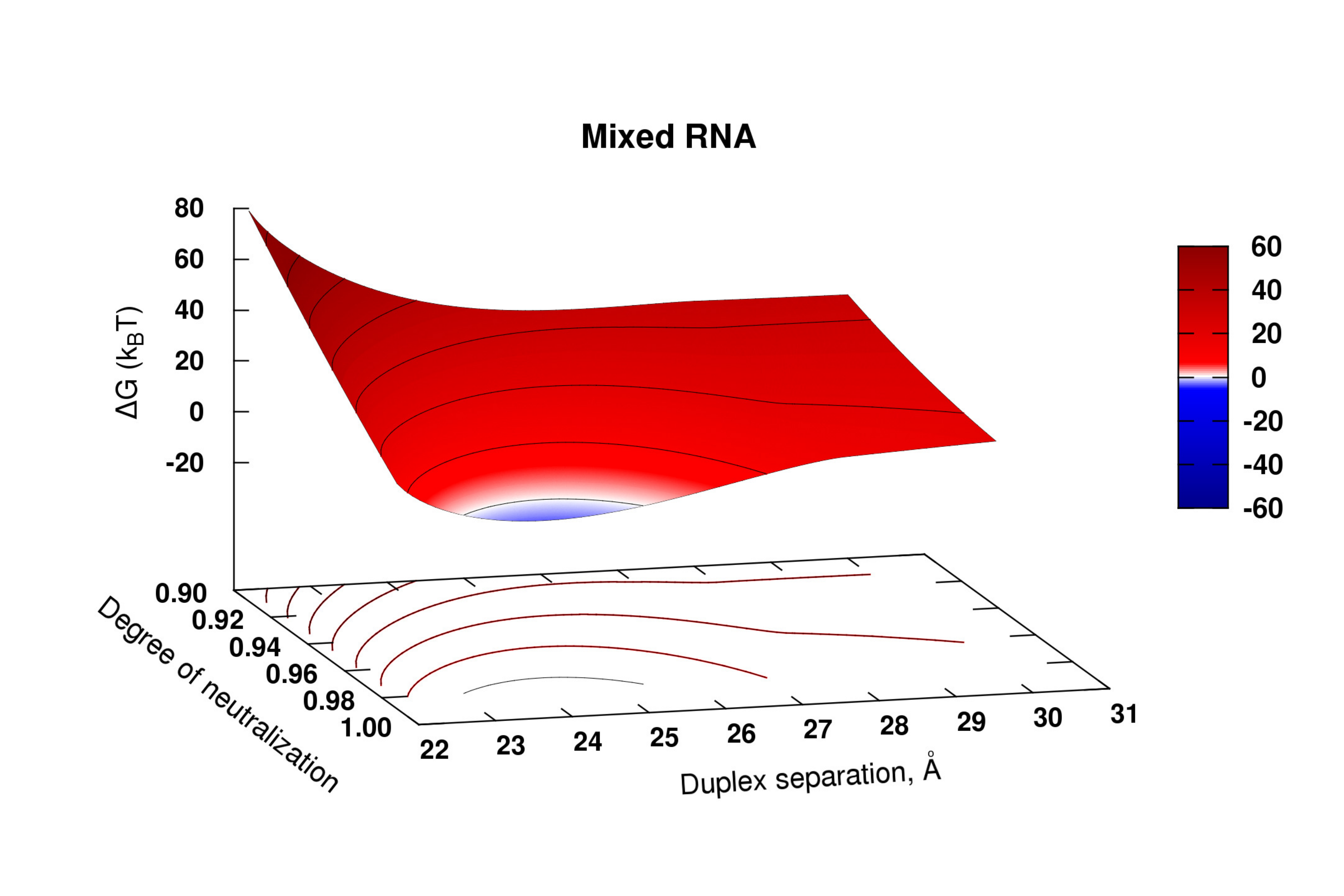}
\caption{Predicted aggregation free energy $\Delta G_{aggr}$ of 
25-bp long DNA and RNA duplexes as a function of the degree of duplex 
neutralization $\Theta$ and the inter-axial duplex separation, $d$.
Condensation conditions, $\Delta G_{aggr} < 0 $, correspond to blue regions
of the parameter space,
while the white band designates $\Delta G_{aggr} \approx 0$. Lines of
equal $\Delta G_{aggr}$ are projected onto the $d - \Theta$ plane.  }
\label{fig:dG3D}
\end{center}
\end{figure}

A more detailed investigation of the model predictions over a range of 
duplex-duplex separation distances and degrees of duplex neutralization 
is presented in Fig.~\ref{fig:dG3D}; the trends 
are exemplified for the mixed sequence DNA and RNA molecules.
\NT{
The aggregation free energy values are estimated using Eqs.~\ref{eq:entrop},
\ref{eq:dGattr3} and \ref{eq:rep3}.
}
Several conclusions can be made that further support the validity 
of our approach. 
%

	For both DNA and RNA, a region of $\Delta G_{aggr} < 0$ as a function 
of inter-duplex separation $d$ in the aggregated phase exists at large enough 
degrees of duplex neutralization $\Theta$ by CoHex ions.
For the mixed DNA duplex, the region of the favorable condensation conditions
begins at $\Theta > 0.92$ with a minimum of $\Delta G_{aggr}(d)$
at the inter-duplex separation $d=26$ \AA. This minimum is not 
too different from the known experimental value $d=28$ \AA\ measured
for long DNA strands \cite{RauParsegian1992,Qiu2013}. With the increase
of $\Theta$, the minimum slightly shifts toward the smaller separations.

	The existence of the minimum of $\Delta G_{aggr}(d)$ can be 
rationalized by considering
differences in scaling behavior of the attractive and repulsive components 
of $\Delta G_{aggr}$ within our model. 
The attractive part $\Delta G_{attr}$ scales linearly
with the number of CoHex ions
in the shells overlap $\Delta N_s$ (Eq.~\ref{eq:dGattr}),
which increases as the inter-duplex separation $d$ decreases.
At the same time, the repulsive contribution $\Delta G_{el-rep}$
is quadratic in $\Delta N_s$, Eqs.~\ref{eq:eta} and \ref{eq:correction}.

	Also, it is reassuring that at typical DNA condensation conditions,
the predicted value of $\Delta G_{aggr}$ per base pair is a fraction of $k_BT$, 
in agreement with the experimental estimates \cite{Bloomfield1997}.

Relative to the DNA, the region of the parameter space ($d$,$\Theta$) where 
the RNA duplexes may be expected to aggregate is very narrow, suggesting
that almost complete neutralization of the RNA duplex by bound
CoHex ions is required prior to aggregation. This result is 
consistent with the notion that RNA is generally
harder to condense than the equivalent in sequence DNA \cite{Li2011,Tolokh2014},
requiring a much larger CoHex concentrations in the solution phase.

Several testable predictions can be made directly from Fig.~\ref{fig:dG3D}.
A close examination reveals that, compared to the DNA,
the aggregated phase of RNA is expected to have shorter inter-duplex distances,
$\sim$1.5 \AA\ shorter than for the aggregated DNA phase.
Physically, this is because even if the RNA neutralization were 100\%,
where $\Delta G_{el-rep} \approx 0$, a relatively weak
attractive force at 28 \AA\ of inter-duplex separation
would not be enough to overcome the entropic cost of aggregation
of short 25-bp RNA duplexes, $\sim$22 $k_BT$ per duplex.
At inter-duplex distances shorter than 28 \AA, an additional
attractive contribution comes into play, the one resulting from
the counterions in the more populated ``internal" ion binding shell of the RNA.
These shells begin to overlap with the ``external" ion binding shells of 
the opposite duplexes at the separation $d < 28$ \AA, Fig.~\ref{fig:overlap2}.
That additional attractive force tips the total attraction-repulsion 
balance towards RNA condensation
at 25 \AA\ inter-duplex separation and near 100\% ($\Theta=0.99$) duplex neutralization.
However, as the duplex-duplex separation decreases further in the already tightly
packed aggregate, both the configurational entropy loss and
the electrostatic repulsion increase, quickly drive the
RNA out of the condensation regime.
 
	Another specific prediction that follows directly from 
the free energy balance within our model is
that longer duplexes are expected to condense better than shorter ones. 
The explanation, based on
Eqs.~\ref{dGagr},\ref{eq:dGattr},\ref{eq:correction},\ref{eq:entrop}, is 
as follows.  All of the contributions to $\Delta G_{aggr}$, except 
the configurational entropy loss, Eq.~\ref{eq:entrop}, 
are proportional to the duplex length $H$ 
(if one neglect the end effects, which is a reasonable approximation 
for the number of base pairs in a duplex $\gg 1$). In contrast,
the unfavorable configurational entropy change 
scales logarithmically with $H$, Eq.~\ref{eq:entrop}.
Thus, as $H$ increases, the relative (per unit length) contribution of 
the entropic term to $\Delta G_{aggr}$ diminishes. 
Since the entropic term resists condensation, 
we expect longer duplexes to condense better.

\subsection{Over-all model performance compared to experiment}

It is reassuring that our estimations based on MD simulation results for 
CoHex ion distributions produce CoHex binding affinities of the same magnitude 
as earlier estimates of CoHex-DNA binding free energies based on equilibrium dialysis 
and single-molecule magnetic tweezers study \cite{Tod2008} (-14.5 $k_BT$),
isothermal titration calorimetry data \cite{Matulis2000} (-13.7 $k_BT$),
and $^{59}$Co chemical shift NMR measurements \cite{Braunlin1987} (-8.6 $k_BT$).
The predicted CoHex affinities to the ``external" ion binding shells of 
the nearly neutralized DNA duplexes are smaller than the above experimental estimates, 
which makes sense since the latter were 
derived from the apparent binding constants in the limit of zero binding; i.e., 
in the absence of the mutual repulsion between bound CoHex ions that decreases
the effective affinity in our calculations.

More importantly, the values in Table~\ref{tbl:28A} correctly predict
the aggregation propensity of different nucleic acid duplexes,
ranging from the favorable aggregation energies for the poly(dA):poly(dT) and
mixed sequence DNA constructs to the unfavorable energies for the DNA:RNA hybrid
and RNA duplex at the same degree of duplex neutralization.
The fact that the order of the condensation propensities observed in the condensation
experiments \cite{Tolokh2014} are captured
by our semi-quantitative model, last line in Table~\ref{tbl:28A},
provides further support
for the robustness of the conceptual picture we have developed.

\NT{
We have verified that the results for the aggregation free energy and its
components presented above are robust with respect to simulation details:
water model used for simulations, initial conditions and details of 
the RNA duplex simulation protocol (see Supplemental Material, 
``Robustness to force-field details  and initial conditions").
}

\section{Methodological details}

\subsection{Additional assumptions and considerations.} 

\begin{enumerate}

\item Since the condensation experiments \cite{Tolokh2014} we are trying to explain
have been conducted at low (20 mM) concentrations of monovalent ions in solution,
when they produce a negligible effect on the bound CoHex ions \cite{Braunlin1987},
the effects of monovalent salt are omitted from our semi-quantitative model.
Whenever appropriate, we will discuss the consequences of neglecting
the presence of small amount of monovalent salt.

\item
The fractions of NA duplex charge neutralized by bound CoHex ions, $\Theta$,
in the solution and aggregated phases are considered to be the same.
\NT{ 
That is the aggregation energy we estimate does not include 
the contributions related the changes of CoHex concentration
(and $\Theta$) when CoHex ions are added to the solution phase
prior to the aggregation and replace the monovalent counterion
atmosphere around the NA duplexes.
}


\item
The aggregated phase as a whole is assumed to be net neutral.
\NT{In addition to bound CoHex ions, a small amount of unbound ions between
the duplexes in a bundle or between the layers of duplex bundles are assumed
in the aggregated phase.
}
This assumption of the overall neutrality of the aggregated phase
allow one to neglect the interactions between the duplexes beyond
the nearest neighbors.

\item 
We use the degree of NA duplex neutralization $\Theta$ as a convenient proxy 
for bulk CoHex concentration.
Even though CoHex concentration in the bulk is the quantity most relevant for 
the thermodynamics analysis presented, it is not easily accessible experimentally.
Typically, NA condensation experiments simply report the total CoHex concentration 
in the solution prior to the condensation, which does not equal 
the bulk CoHex concentration. The relationship between the two and $\Theta$ is
complex \cite{Shklovskii1999a,Nguyen2000}, but it is monotonic,
which justifies the use of $\Theta$ for our purposes: we are interested
in the relative NA condensation propensities, not their absolute values.
The relation between $\Theta$ and the bulk multivalent ion concentration has 
some non-trivial form \cite{Shklovskii1999a,Nguyen2000} different from 
the original Manning predictions \cite{Manning1969,Manning1978} and predictions
based on the Poisson-Boltzmann (PB) model \cite{Ramanathan1983,Stigter1995,Deserno2000,OShaughnessy2005}
due to correlations between bound multivalent ion at the surface of NA molecules.
We do not need to consider the exact relation here. 
The only general property of the layer of bound (condensed) trivalent counterions 
at the NA duplex surface we will use is that the degree of NA duplex
charge neutralization by these counterions, $\Theta$, weakly depends
on the bulk ion concentration and monotonically increases with the increase 
of the latter \cite{Shklovskii1999a,Nguyen2000}. At low (mM) trivalent 
ion bulk concentration the value of $\Theta$ is close to 
the Manning theory prediction for the double-stranded DNA, $\Theta \approx 0.9$.

\end{enumerate}

\subsection{Estimation of CoHex binding affinity $\mu_a$.}

We approximate the average binding energy that a CoHex ion bound in the ``external"
binding shell of NA duplex 1 gains when it enters the ``external" binding shell of
the (neighboring) NA duplex 2, by the CoHex binding affinity $\mu_a$
to the ``external" ion binding shell of an isolated duplex in the solution phase
at near neutralizing conditions.  This approximation
is based on the assumption that minimal changes in the CoHex distributions occur
when the two NA duplexes approach each other at the inter-duplex
distances relevant to NA condensation.

The above assumption was tested in Ref.~\cite{Tolokh2014} for  homopolymeric
poly(dA):poly(dT) DNA and mixed sequence RNA duplexes.
By comparing the superposition of two independent single duplex CoHex distributions
with CoHex distribution around a pair of NA duplexes separated by 26 \AA,
no changes in the number of ``internally" bound
(inside 12 \AA\ from the helical axis) CoHex ions were observed.
A small increase (1.1 ions) in the ``external" ion binding shell
(12 to 16 \AA\ from the helical axis) of a duplex
in both DNA and RNA duplex pairs was seen. The latter was suggested to
be due to some redistribution of the bound ions in the external shell toward
the shell overlapping region of the pair, which can be neglected in the case
of six neighbors of the hexagonal packing considered in this work.

CoHex binding affinity $\mu_a$ to the ``external" ion binding shell of NA duplex
can defined as a difference of the excess chemical potentials of CoHex ion
in the ``external" shell of the duplex, $\mu^s_{excess}$, and in the bulk, $\mu^b_{excess}$,
\begin{equation}
\mu_a = \mu^s_{excess} - \mu^b_{excess}  \, .
\label{mua_def}
\end{equation}
The quantity reflects the change of the ion
interaction with its environment as the ion moves from the bulk towards
the surface of charged NA molecule.

The chemical potential of CoHex ion, $\mu$, can be written as
\begin{equation}
\mu = \mu_{excess}  + \mu_{ideal} \, ,
\label{cp}
\end{equation}
where
\begin{equation}
\mu_{ideal}= k_BT \mathrm{ln }(\rho_n C) 
\label{id}
\end{equation}
is a portion of the potential
that can be treated as a chemical potential of an ideal gas with
a particle number density $\rho_n$ corresponding to the CoHex density
\cite{Widom2002},
\NT{
$C$ is a constant which does not depend on $\rho_n$.
}
At equilibrium between the bulk ($b$) and the layer of bound CoHex ions in
the ``external" ion binding shell ($s$),
$\mu$ is constant throughout the system:
\begin{equation}
\mu^s_{excess}  + \mu^s_{ideal} = \mu^b_{excess}  + \mu^b_{ideal}
\label{eq:constntmu}
\end{equation}

From Eqs.~\ref{id} and \ref{eq:constntmu}  and
the definition of the ion binding affinity, Eq.~\ref{mua_def}, we arrive
at Eq.~\ref{mua}, i.e. $\mu_a = - k_B T\ln \left(\rho_s/\rho_b\right)$.

CoHex number density in the ``external" ion binding shell
$\rho_s = N_s^0 / V_s$ is estimated using
the average number of CoHex ions in this shell $N_s^0$ determined
from the MD simulations \cite{Tolokh2014}
and presented in Table~\ref{tbl:NAR}.
With the height $H=88$ \AA\ for the two DNA duplexes and $H=76$ \AA\ for
the DNA:RNA hybrid and RNA duplexes, the ``external" shell volume,
$V_s=\pi(r_s^2 - r_i^2)H$,
constitutes 30964 and 26741 \AA$^3$, respectively.

\begin{table}[h!]
\NT{
\caption{Average numbers of bound CoHex ions in the ``external" and ``internal" 
ion binding shells around the four NA duplexes estimated from the MD simulations 
\cite{Tolokh2014}.} 
\label{tbl:NAR}%
\centering
\small
\begin{tabular}{@{\vrule height 10.5pt depth4pt  width0pt}|l|c|c|c|c|}
\hline
                          & DNA(dA:dT) &   DNA   &   HYB    &    RNA \\
\hline
``External" shell ions, 
               $N_s^0$    &   9.8       &    8.5  &   3.3    &    2.0  \\
\hline
``Internal" shell ions,
               $N_i^0$    &   4.6       &    4.2  &   7.7    &    9.4 \\
\hline
All bound CoHex ions, 
                 $N_0$    &  14.5       &   14.4  &  14.1     &  14.7 \\
\hline
 \end{tabular}
}
\end{table}

\subsection{Estimation of repulsive interaction $\Delta G_{el-na}$ between the two 
unscreened NA duplexes.}

We define $\Delta G_{el-na}(d)$ as the difference of the electrostatic free energy
of the two duplexes in water, $\Delta G_{el}$, at the separations $d$ and $\infty$.
The $\Delta G_{el}$ is estimated as \cite{Onufriev2010}
\begin{equation}
 \Delta G_{el}=\Delta G_{solv} + E_{Coul} \, .
\label{eq:el}
\end{equation}
Here $\Delta G_{solv}$ is the solvation energy of the two duplexes,
estimated from the solution of the PE, and
$E_{Coul}$ is the Coulomb charge-charge interactions in the system
computed with AMBER12 \cite{amber12}. 
\NT{
Both terms in Eq.~\ref{eq:el} are computed using charge distributions
on NA duplexes determined from the AMBER ff99bsc0 nucleic acid force-field
\cite{Cheatham1999,Perez2007}.
}
The $\Delta G_{el}$ for the two infinitely separated duplexes is calculated
as the sum of the electrostatic free energies of the two isolated molecules.

We estimate $\Delta G_{el}$ by solving the Poisson equation (PE) for
the two parallel unscreened duplexes and averaging the interaction energies
over 12 different mutual duplex orientations.
The interactions are computed at two different inter-axial
separations, $d=24$ and $28$ \AA. The orientations were changed by rotating
one of the duplexes around its helical axis with $30^\circ$ increment.
The interactions at other inter-axial duplex separations were estimated using
an interpolation, see below. 

The PE was solved using MEAD solver \cite{Bashford1997mead}
with three levels of focusing and with 281 grid points in each dimension: 
the coarsest grid spacing was 5.0 \AA, and the finest 0.5 \AA.
The internal (NA duplex) and the external (water) dielectric constants
were 4.0 and 78.5, respectively \cite{Jayaram1989,Stigter1998}.

The interaction $\Delta G_{el-na}(d)$  between the two parallel NA duplexes in water
is then estimated as the difference of the orientationally averaged
$\Delta G_{el}$ at a separation $d$ and $\infty$,
\begin{equation}
 \Delta G_{el-na}(d) = \langle \Delta G_{el}(d) \rangle - \Delta G_{el}(\infty) \, .
 \label{eq:eltot}
\end{equation}

\NT{
The value of $\Delta G_{el-na}(d)$ at the inter-duplex separations different
from the two distances ($d_1=24$ \AA\ and $d_2=28$ \AA) for which it was directly
calculated is then estimated using the logarithmic interpolation,
\begin{eqnarray}
\Delta G_{el-na}(d) = \Delta G_{el-na}(d_1) +  \hspace{3cm} \nonumber \\
+ \left(\Delta G_{el-na}(d_2)-\Delta G_{el-na}(d_1)\right) \frac{\ln(d_1/d)}{\ln(d_1/d_2)} .
\label{eq:eltot_d}
\end{eqnarray}
}

At high degree of duplex neutralization, $\sim$90\%, the variation
of the scaled $\Delta G_{el-na}(d_2)$ 
with one of the duplex rotating about its
helical axis does not exceed 0.3 $k_{B}T$, consistent with an earlier estimate
of $\sim 0.5 k_{B}T$ per 25 base pairs based on a different model
\cite{Kornyshev2013}.
The insignificance of the variation justifies the use of the simple averaging
for $\Delta G_{el-na}$.
Small ($\Delta Z/H \ll 1$) axial translations of one duplex with respect to another
reduce the repulsive contribution mostly through the reduction of $\Lambda$
which is proportional to the shell overlapping volume ($\Lambda \sim \Delta N_s \sim \Delta V_s$).
At the same time, the change in $\Delta N_s$ leads to a reduction of the attractive contribution
to the aggregation energy resulting in a destabilization of the duplex bundle.

\subsection{Atomistic simulations.}

The details of CoHex ion distributions and its chemical potential 
were extracted from recently published
MD simulations \cite{Tolokh2014} of the same four NA duplexes
considered here. A brief description of the protocols used
are presented below. Other details of the duplex structure preparations and simulations
are described in \cite{Tolokh2014}.

All-atom MD simulations were used to generate CoHex ions distributions
around four 25-bp NA duplexes: homopolymeric poly(dA):poly(dT) DNA,
mixed sequence DNA, DNA:RNA hybrid and RNA.
Each system contained one 25-bp NA duplex, a neutralizing amount of 16 CoHex ions
and 16880 TIP3P \cite{TIP3P} water molecules. 
To avoid uncertainties associated with monovalent ions force field parameters
\cite{Yoo2012a}, no monovalent salt was added to the simulated systems.
The simulated trajectories were 300-380 ns long, generated at 300 K temperature in 
the canonical (NVT) ensemble. The simulations were performed
using AMBER12 package \cite{amber} and ff99bsc0 force-field \cite{Cheatham1999,Perez2007}.
Since no measurable effects of CoHex on the DNA duplex structures were experimentally
observed \cite{Tolokh2014}, the homopolymeric and mixed sequence DNA duplexes were
restrained to their B$^{\prime}$ \cite{Alexeev1987} and canonical B-form,
respectively, during the simulations. These restraints minimized possible structure
bias due to the use of imperfect modern force-fields \cite{Mackerell2004}.
On the other hand, the addition of CoHex has lead to the experimentally observed
changes in RNA helical structure \cite{Tolokh2014}.
Therefore, the RNA and DNA:RNA hybrid duplexes were simulated unrestrained
allowing the duplex structures to relax when CoHex ions bind to these molecules.
The CoHex ion distributions were analyzed using 28,000, 30,000 and 34,000 
snapshots extracted from DNAs, DNA:RNA hybrid and RNA duplex trajectories, respectively.

\section{CONCLUSIONS}

Counterion-induced condensation of nucleic acids (NA) is a complex phenomenon 
that attracted experimental and theoretical attention for several decades. 
Recent experiments in which trivalent CoHex ions failed to condense short 25-bp 
double-stranded (ds) RNA fragments, in contrast to equivalent dsDNA fragments, 
demonstrated that our understanding of the process needs further refinement. 
The fact that dsRNA resits condensation at the same conditions where dsDNA readily 
condenses is counter-intuitive: indeed, the binding of CoHex ions to dsRNA is 
stronger than to the DNA, and the two highly charged molecules are virtually 
the same at the level of ``charged rods" model. As it turns out, however, details 
of counterion distributions around dsDNA and dsRNA are very different with respect 
to proximity of bound CoHex to the helical axis. Namely, 
the NA-NA attraction, and, as a consequence, 
condensation propensity, is determined mainly by the fraction
of counterions bound to the external (outermost) surface of
the double-helix.

Here, we have developed the first semi-quantitative model of nucleic acid 
aggregation (condensation) induced by trivalent CoHex (Co(NH$_{3})_{6}^{+3}$)  
counterions in which the radial distribution of the bound counterions
is the key ingredient.  Namely, the counterions bound to a NA duplex
are partitioned into an ``external" and ``internal" ion binding shells; 
the fraction of ions in each shell, and their affinity to the nearly neutralized 
duplex can be accurately quantified from converged ion distributions obtained 
from all-atom molecular dynamics simulations in explicit solvent. 
Within the model, the source of the short-range attraction between
two approaching NA duplexes are the oppositely charged CoHex ions from 
the overlapping region of the ion binding shells of these duplexes. 
Basic thermodynamic arguments are then used to estimate the aggregation 
free energy components 
corresponding to CoHex-mediated attraction between the duplexes and 
the opposing residual repulsion of the nearly neutralized duplexes and 
configurational entropy loss upon aggregation of the short duplexes.
Importantly, robustness of the model estimates to simulation details has been 
thoroughly verified. 

The key conclusion from this study is that our semi-quantitative model based 
on the ``ion binding shells'' framework is able to reproduce the correct order 
of condensation propensities seen in experiment for various NA duplexes, 
some of which differ in fairly subtle way, e.g., poly(dA):poly(dT) DNA 
vs.~mixed sequence DNA.  
The key role of the fraction of multivalent ions bound in the ``external" 
ion binding shell of NA duplexes in nucleic acid condensation 
is clarified. The larger the fraction of the externally bound multivalent
ions, the larger the condensation propensity of a nucleic acid molecule.

Another conclusion is 
a relatively small value of the attractive interactions between the RNA duplexes 
at the inter-axial distances and degree of neutralization
at which DNA duplexes begin to aggregate. To overcome the relatively large 
entropic cost of the aggregation of short RNA duplexes 
it is necessary both to increase the attractive component and 
to minimize the electrostatic repulsive contribution by a more complete
duplex neutralization compare to the case of the DNA.
We therefore predict that RNA condensation occurs at inter-helical distances 
smaller than ones experimentally found in DNA aggregates. Consistent 
with experiment, the model predicts that higher degree of RNA neutralization 
is needed as well (in experiment, RNA condensation starts at considerably higher
CoHex concentrations than for the equivalent DNA molecules). 

We look forward to experimental testing of these predictions through anomalous small-angle X-ray 
scattering and related approaches. 

In contrast to some of the available models of counterion-induced NA condensation, 
ours is simple enough to allow for 
analytical estimates and analyses of trends. 
For example, we show why longer NA 
fragments are expected to condense easier than shorter ones. At the same 
time, the apparent simplicity of the model does not come at the expense of 
its physical realism: the key component of the model -- counterion affinity 
to the ``external" ion binding shell and the fraction of ions in this shell
are computed from explicit solvent atomistic simulations, which is arguably 
the most accurate practical approach to such estimates to-date. 
The details of the NA duplex geometry and sequence
are implicitly included in the values of these parameters.  
The atomistic level of details ``built into'' 
the basics physics model greatly enhances versatility of the latter.    
In particular, in the future we are interested in exploring a wider range 
of multivalent ion types, and look more closely into sequence dependence of 
NA condensation.

\clearpage

\begin{acknowledgments}
This work was supported by the National Institutes of Health (NIH) R01 GM099450.

\end{acknowledgments}

\bibliographystyle{pnas}
\bibliography{naf}

\end{document}